\documentstyle[12pt,titlepage,fleqn]{article}

\newcommand{\be}{\begin{equation}}
\newcommand{\ee}{\end{equation}}
\newcommand{\ba}{\begin{eqnarray}}
\newcommand{\ea}{\end{eqnarray}}
\newcommand{\om}{\omega}
\newcommand{\q}{\hat{q}}
\newcommand{\p}{\hat{p}}
\newcommand{\cO}{{\cal O}}
\newcommand{\non}{\nonumber}

\newcommand{\oh}{\frac{1}{2}}

\begin{document}

\begin{titlepage}
\vspace*{0.1truecm}
\begin{center}
\vskip 1.0cm
{\large\bf  CAN PLANE WAVE MODES BE PHYSICAL MODES IN SOLITON MODELS?}

\vskip 2.0cm

{\large F. ALDABE\footnote{E-mail: faldabe@phys.ualberta.ca}}

{\large\em International Center for Theoretical Physics\\
P.O. Box 586, 34100 Trieste, Italy}\\
\vspace{.5in}
\today \\
\vspace{.5in}
{\bf ABSTRACT}\\
\begin{quotation}
\noindent

I show that plane waves may not be used as asymptotic states in soliton
models because they describe unphysical states.
When asymptotic states are taken to be physical there is no T-matrix of
$\cO(1)$.

\end{quotation}
\end{center}

\end{titlepage}

It has been argued by some authors \cite{ohta}\cite{uehara}
that soliton models have Born terms
of $\cO(1)$. Similar statements where made more recently by
the authors of \cite{mattis} for the Skyrme model.
The existence of these terms implies that these models have a Yukawa
coupling between the fluctuations, the mesons of the theory, and the
soliton, the baryon of the theory.  There are no  Born terms due to
Heitler-Bhabha mechanism since there are no
linear terms in the fluctuations.  These terms which represent
the Yukawa coupling are absent because the soliton is defined by minimizing
the equations of motion.  It is thus impossible to have linear terms
in the fluctuations which are of $\cO(\sqrt{M})$ where $M$ is the mass
of the soliton.  Rather,
the surprising result leading to the confirmation of a Yukawa coupling
 is obtained after treating
the zero modes of the theory
and making use of plane waves as asymptotic states.
It is the purpose of this paper to show with simple arguments that the
asymptotic states used in the calculation of the above authors are unphysical,
to show that there are no Born terms and thus no Yukawa coupling in this
theory.

I will make use
of the Dirac treatment of collective coordinates.
Let me recall how this treatment is used in the case of solitons
in $1+1$ spacetime dimensions.  The reader is referred to \cite{tom}
for a detailed discussion.

We begin by solving the classical equations of motions.
There will be two solutions in the $\phi^4$ model and there will be infinite
solutions
in the sine-Gordon model.  These solutions will be classified by their
topological charge.  The one baryon solution will have topological
charge $1$.  It will not be invariant under translations, its space derivative
must be non vanishing in order to have non vanishing topological charge.
However, the Hamiltonian will be invariant under space translations.  If
one examines the Hamiltonian, she or he will find a zero mode.  It will
lead to infrared divergences.  In order to integrate this zero mode,
one usually introduces collective  coordinates.
A constraint $f$ follows from the definition of collective momenta
\be
f=J_t-\Pi\label{cc1}
\ee
where $J_t$ is the generator of the translations and $\Pi$ is the collective
momenta conjugate to the collective coordinate $X$.
Physical magnitudes should commute with the constraint while
physical states must be annihilated by its action \cite{bk}.
Upon considering fluctuations of the field, the constraint
will take the form
\be
f=\int\phi_c'\p+\int\q'\p-\Pi.\label{tombou}
\ee
Here, $\phi_c$, the classical solution of the equations of motion is
of $\cO(M^{\frac{1}{2}})$, $M$ being the mass of the soliton, while $\q$,
the fluctuation about the classical solution is of $\cO(1)$. $\p$ is
the conjugate of the fluctuation
\be
i[{\hat{p}}(x),\hat{q}(y)]=\delta(x-y).\label{poi}
\ee

In order to take into account the dynamical
information of the constraint (\ref{tombou}), Gervais and
collaborators \cite{gjs} and Tomboulis \cite{tom} have
introduced a transformation which acts on the fluctuations as
\ba
\p&\rightarrow&\p+\frac{\Pi-\int\p\q'}{M+\int\phi'_c\q'},\non\\
\q&\rightarrow&\q\non\\
\Pi&\rightarrow&\Pi\non\\
X&\rightarrow&X
\ea
in order to have simple expressions for the Dirac commutator and thus include
the effect of the constraint.
\footnote{
Here canonicity is lost because it implies that
\be
i[{\hat{p}}(x),\hat{q}(y)]\ne\delta(x-y).
\ee
The effects of this non canonical transformation have been evaluated in
the sine-Gordon model in \cite{yo} by comparing with the BRST treatment
which does not make use of such a transformation.
It has been shown not to affect physical
quantities to ${\cal O}(M^{-1})$.}

This transformation has the purpose of taking the constraint to the form
\be
f=\int\phi_c'\p.
\ee
It also acts on the Hamiltonian, however its only effect is to
add to it terms which can be treated perturbatively because they are
of $\cO(M^{-\oh})$ or inferior.  Thus, the $\cO(1)$
quadratic Hamiltonian remains
invariant under such a transformation.

So far we have enlarged the original phase space containing the dynamical
variable
$\phi$ and its conjugate to a new phase space which has the dynamical variable
$\phi,\;X$, and their conjugates.
In order to recover the original system in this enlarged phase space,
Tomboulis showed for our case, following the ideas of Dirac \cite{sm},
that it is necessary to introduce a gauge condition,
in particular we may choose it to be
\be
G=\frac{1}{M}\int \phi_c'\q,
\ee
as it is usually done,
\footnote{ A choice of gauge $G=\int h\q$, with $h$ an arbitrary function of
space,
 will lead to the same results as
long as $[G,f]\ne 0$ because physical magnitudes, those commuting with
the constraint $f$ are independent of the choice of gauge.  Furthermore,
a change of gauge which preserves its order of magnitude with respect to $M$
cannot modify the perturbative expansion in $M$.}
and then, the second class constraint, along with this gauge yield the Dirac
commutator which reads in terms of the Poisson commutators
\be
[A,B]_D=[A,B]_P-[A,f]_P\frac{1}{[G,f]_P}[G,B]_P
+[B,f]_P\frac{1}{[G,f]_P}[G,A]_P,
\ee
where the subscript D (P) refers to  Dirac (Poisson) commutator.
As it is well known, this commutator will yield the equations of motion
for all dynamical variables in the enlarged phase space.

Thus, a simple calculation shows using  (\ref{poi}) that
\be
i[{\hat{p}}(x),\hat{q}(y)]_D=\delta(x-y)-\frac{\phi_c'(x)\phi_c'(y)}{M}.
\label{idiota}
\ee

Suppose that we would like to calculate the S-matrix starting with asymptotic
states which are plane waves. We would split the quadratic Hamiltonian as
follows
\ba
H^{(2)}&=&H^{(2)}_o+V^{(2)}\\
H^{(2)}_o&=&\int(\oh\p^2+\oh\q^{'2}+\oh \mu^2\q^2)\label{h0}\\
V^{(2)}&=&\int (V_2-\mu^2)\q^2
\ea
where $V_2=\frac{\partial^2V}{\partial\phi^2}|_{\phi_c}$ and $\mu$ is the mass
of the asymptotic fluctuations, and diagonalize $H^{(2)}_o$ by expanding
the fluctuations in terms of plane wave modes
\ba
\q&=&\sum_k \frac{i}{\sqrt{2\om_k}}(\psi_k a_k-\psi_k^*a^+_k)\non\\
\p&=&\sum_k \sqrt{\frac{\om_k}{2}}(\psi_k a_k+\psi_k^*a^+_k)
\ea
where $\psi_k$ are solutions to the plane wave equation, in the hope
that as shown in \cite{ohta} and \cite{uehara} the interaction $V^{(2)}$ will
yield
Born terms.

The zeroth order Hamiltonian would be
\be
H^{(2)}_o=\sum_k\om_k(a^+_ka_k+a_ka_k^+).
\ee
The intrinsic vacuum would be defined by the condition
\be
a_k|vac>=0,
\ee
for any $k$.
\footnote{The collective coordinate and its conjugate do not act
on the intrinsic vacuum since the basis which we have chosen to perform
our perturbative expansion is written as
\be
|collective \;vacuum>\otimes|vac>.
\ee
This decomposition is possible because the Hamiltonian in the enlarged
phase space will have as leading term for the collective sector
the term
\be
\frac{\Pi^2}{2M}.
\ee
Thus in the low energy regime, the soliton can be treated
as a galilean particle.}

 In order to trusts that the results are correct
we should verify that the states we use are physical.
As is well known,
this implies that the states must be annihilated by the constraint.
For our particular case,
 the in-state $|k,in>$ would be defined as
\be
|k,in>=a_k^+|vac>
\ee
for arbitrary $k$.  This state can be written as
\be
\int \psi_k(i\sqrt{\frac{\om_k}{2}}\q+(\frac{1}{\sqrt{2\om_k}}\p)|vac>.
\label{state}
\ee
Of course, we assume that the vacuum is physical and therefore, it is
annihilated by the constraint.  Then the action of the
constraint on state (\ref{state}) will be
\be
f|k,in>=\sqrt{\frac{\om_k}{2}}\int\phi_c'\psi_k|vac>\label{annh}
\ee
Clearly, it does not vanish for all $k$.  In fact, explicit calculations
in the $\phi^4$ model, where $\phi_c$ is known
 shows it does not vanish for any $k$.  In this model $\phi_c\sim tanh(x)$.
The careful reader might have noticed that in deriving expression (\ref{annh})
I have made use of the Poisson commutator instead of the Dirac commutator.
It is the first commutator which must be used in (\ref{annh}) in order to
verify that the asymptotic state are indeed unphysical.  The reader is referred
to  \cite{bkk} for a detailed explanation of why this is so.

Thus we find a surprising result: plane waves are unphysical because they are
not annihilated by the constraint.  These states carry components of zero
modes.
Thus, we may not trust the results of Ohta \cite{ohta} and Uehara et. al.
\cite{uehara} and their claim that the proper treatment of zero modes lead to
Born terms because they have made their calculations using plane waves
which I have shown are not physical.  In order to calculate the Born terms
one must make use of
states which are physical, that are annihilated by the constraint.
The procedure to follow is to find the normal modes
which diagonalize the zeroth order
quadratic Hamiltonian using the Dirac brackets rather than the Poisson
brackets.

Let us study the main properties of the normal modes which follow
from the use of the Dirac commutator (\ref{idiota})
and the Hamiltonian (\ref{h0}).
A little algebra using (\ref{idiota}) and (\ref{h0})
lead to an equation of motion

\be
\ddot{\q}-\q''+\phi_c'\int\phi_c'\q''/M +\mu^2 (\q-\phi_c'\int\phi_c'\q/M)=0.
\label{fu1}
\ee
Now let us expand $\q$ in prospective normal modes
\be
\q=\sum_l \frac{i}{\sqrt{2\om_l}}(\psi_l a_le^{i\om_l
t}-\psi_l^*a^+_le^{-i\om_lt})\label{nexp}.
\ee
Note that 1) $\phi_c'$ is not a solution of zero energy. 2)
Multiplication of (\ref{fu1}) by $\phi_c'$ and integration imply that
all $\psi_l$ are orthogonal to the zero mode,
insuring that they are  physical because they satisfy
\be
f|l,in>=\sqrt{\frac{\om_l}{2}}\int\phi_c'\psi_l|vac>=0.
\ee
3) multiplication by $\psi_{l'}$
and integration imply that the normal modes form an orthogonal basis which
by 2) is orthogonal to the zero mode.  4) (\ref{nexp}) is invariant
under complex conjugation, thus if $\psi_l$ is solution then $\psi_l^*$ is
also a solution. 5) Since the energy of the soliton is localized and
by the virial theorem is
proportional to $\phi_c'^2$ it follows that far away from the soliton, where
the beam of fluctuations is created, $\phi_c'$ vanishes
and the equation of motion (\ref{nexp})
reduces to the known Klein-Gordon equation, implying that the
solutions in this region will behave like plane waves.

The next step is to calculate the T-matrix to zeroth order.
For this we note that the T-matrix, responsible for the phase shifts,
  for plane waves is given by \cite{ohta}
\be
T(k)=\frac{\om_k^2|\int\phi'_c\psi_k|^2}{M}.\label{tma}
\ee
Expression (\ref{tma}) is nonvanishing for plane waves, but after replacing
the plane waves for physical asymptotic states $\psi_l$
we find a vanishing T-matrix
of $\cO(1)$ due to property 2).
The authors in \cite{mattis} made use of the constraints in the Skyrme model
to derive an effective Hamiltonian.  They should also have made use of
these constraints to define the asymptotic states used to
calculate the Skyrme decay amplitude which they claim to be
of $\cO(1)$.

\end{document}